\documentclass[fleqn,twoside,epsfig,psfig]{article}
\usepackage{espcrc2}

\usepackage{graphicx}
\usepackage[figuresright]{rotating}

\usepackage{amssymb}
\usepackage{graphics}
\usepackage{epsfig}

\newcommand{\Vec}[1]{\vec{#1}\hspace{.7mm}}

\def\12{\frac{1}{2}}

\def\ch{{\rm cosh}\hspace{.5mm}}
\def\sh{{\rm sinh}\hspace{.5mm}}

\def\ve{\varepsilon}

\def\ve{\varepsilon}
\def\D{{\cal D}}

\def\r{\rho}
\def\s{\sigma}

\def\wh#1{\widehat{#1}}
\def\wt#1{\widetilde{#1}}

\def\ol#1{\overline{#1}}

\def\d{\partial}
\def\m{\mu}
\def\n{\nu}

\def\e{\epsilon}

\def\be{\begin{equation}}
\def\ee{\end{equation}}
\def\beq{\begin{equation}}
\def\eeq{\end{equation}}
\def\bea{\begin{eqnarray}}
\def\eea{\end{eqnarray}}
\def\beqa{\begin{equation}\begin{array}{l}}
\def\eeqa{\end{array}\end{equation}}

\def\eqn#1{(\ref{#1})}

\def\eqref#1{eq.~(\ref{eq:#1})}

\def\L{{\it\Lambda}}

\def\pa{\partial}

\def\pa{\partial}

\def\nn{\nonumber}

\newcommand{\AmS}{{\protect\the\textfont2
  A\kern-.1667em\lower.5ex\hbox{M}\kern-.125emS}}

\title{Novel Properties of Massive Higher Spin Fields}

\author{S. Deser\address[MCSD]{Department of Physics, Brandeis University \\
        Waltham, MA 02454, USA}}%

\begin{document}

\begin{abstract}
I outline a series of results obtained in collaboration with A.\
Waldron on the properties of massive higher ($s>1$) spin fields in
cosmological, constant curvature, backgrounds and the resulting
unexpected qualitative effects on their degrees of freedom and
unitarity properties.  The dimensional parameter $\L$ extends the
flat space $m$-line to a $(m^2,\L )$ ``phase" plane in which these
novel phenomena unfold. In this light, I discuss a possible
partial resurrection of deSitter supergravity. I will also exhibit
the well-known causality problems of coupling these systems to
gravity and, for complex fields, to electromagnetism,
systematizing some of the occasionally misunderstood obstacles to
interactions, particularly for $s = 3/2$ and 2.

 \vspace{1pc}
\end{abstract}

\maketitle

\section{Introduction}

It is a particular pleasure to report here on work with Andrew
Waldron on properties of higher spin fields, because these systems
have a longtime and broad following here in Brazil. Although I
will not assume universal familiarity with the background, space
will not permit me to delve into any details either. For these, I
refer you to our papers \cite{DeserWaldron} where you will also
find references to guide you through the enormous and often
contradictory earlier literature.  A more detailed review article
should also appear in due course.

Observationally, we seem to live in a universe whose vacuum state
is the deSitter (dS) geometry of constant positive, rather than
flat, curvature. At the same time we know that string theory, when
expanded in terms of local actions at low energy, gives rise to
towers of massive higher spin fields beyond the currently observed
world of the standard model, with spin 0, $\frac{1}{2}$ and 1.
These broad facts immediately force us to go beyond flat space,
where representations of the Poincar\'{e} group simply fall into a
massless/massive dichotomy, the former having 2, and the latter
have 2s+1, degrees of freedom (DoF) in $D$=4.  This is not
entirely unknown territory -- it was initially explored by Dirac,
in the 1930s and the so-called cosmological extensions of $N$=1
SUGRA also seemingly required ``massive" and ``gauge" spin 3/2
particles in Anti-deSitter (AdS, $\L < 0$) gravity.  The first
``modern" results \cite{DeserNepomechi} concerned (linear, in no
way ``graviton-like") massive spin 2 fields, where the unusual
properties we will develop here are already visible:  At ``tuned"
values of the dimensionless ratio $m^2/\L$, novel local gauge
invariances appear and remove the helicity 0 component of the
original model's helicities $(\pm 2,\: \pm 1,0)$.  A systematic
study \cite{Higuchi} of the corresponding 2-point functions then
revealed the presence of a forbidden, negative norm, region of
this $m^2/\L$ ratio.  This turns out to be the tip of a major
iceberg: ``partial masslessness", with associated novel gauge
invariances and unitarily forbidden regions of the $(m^2, \L )$
plane occurs for all fermions and bosons with spin $s>1$.
Important consequences include null propagation at the critical
values, as well as energy stability and unitarity properties of
the corresponding quanta. These may even affect the old question
whether SUGRA is tenable in dS, despite the fact that the spin 3/2
particle's mass parameter is proportional to $\sqrt{-\L}$.

The second set of questions -- consistency difficulties of
coupling massive spin $>$1 fields to gravity (not just in (A)dS!),
and if complex, to electromagnetism, is a very old and actually
well-understood topic particularly for the key, $s=3/2$ and $s=2$,
cases. Our systematization of it is mentioned here because there
are still occasional misunderstandings (even in Brazil) of the
established results.  We use the notion of characteristics to show
that acausality can be seen to arise as the onset of spacelike
propagation at strong enough external magnetic fields, for
example, in addition to the other ills, namely loss of  correct
DoF count, because essential free-field constraints can be negated
by the coupling.


\section{de Sitter Space}

\label{dS_space}

We begin with a brief summary of our geometries: dS$_{n+1}$
spacetime, the coset $SO(n+1,1)/SO(n,1)$, geometrically described
by the one-sheeted hyperboloid
 \begin{eqnarray}
Z^MZ_M & = & -(Z^0)^2+\sum_{i=1}^n(Z^i)^2+(Z^{n+1})^2=1\, ,\nonumber \\
M & = & 0. ... n+1 \; , \label{hyperboloid} \end{eqnarray}
 embedded in $\mathbb R^{({n+1},1)}$. The
cosmological constant $\Lambda$ is positive in~dS space and
usually enters the equation~\eqn{hyperboloid} as $n/\Lambda $ (=1
in our units) on the right hand side. Lack of space precludes
details of the analogous AdS picture, where $SO(n,2)$ is the
isometry group.

The~dS group $SO({n+1},1)$ acts naturally on $\mathbb
R^{({n+1},1)}$ with generators
 \be M_{MN}=i(Z_M\d_N-Z_N\d_M) \ee
obeying the Lie algebra
 \begin{eqnarray}
\lefteqn{[M_{MN},M_{RS}]=i\eta_{NR}M_{MS}-i\eta_{NS}M_{MR}}\nonumber \\
&&+ \; i\eta_{MS}M_{NR}-i\eta_{MR}M_{NS}\, . \label{Lie}
\end{eqnarray}
 Using a
coordinate system in which spatial sections are flat, one can
rewrite this algebra in terms of
 \begin{eqnarray}
 P_i & \equiv & M_{i0}+M_{{n+1},i}\, ,\;\;\; D \; \equiv \; M_{{n+1},0}\,
 ,\nonumber \\
K_i & \equiv & M_{i0}-M_{{n+1},i}\, , \;\;\; M_{ij} \label{gens0}
 \end{eqnarray}
  The latter satisfy the $so(n)$ angular momentum Lie algebra, so
identifying the remaining generators $P_i$ as momenta, $D$ as
dilations and $K_i$ as conformal boosts, their algebra generates
the Euclidean conformal group in $n$ dimensions, with $i, j =
1...n$.

In terms of the embedding coordinates
 \begin{eqnarray}
 \lefteqn{Z^M=(gX)^M=} \nonumber \\
 && \Big(
\sh t+\frac12\, e^t \Vec x^2\, ,\,  e^t\vec x\, , \,  \ch t -
\frac12\, e^t \Vec x^2 \Big)\, , \label{embed}
 \end{eqnarray}
 the interval becomes
 \begin{eqnarray}
 ds^2 & = & dZ^M \eta_{MN} dZ^M \nonumber \\
 & = & -dt^2+e^{2t}d\Vec x^2
\equiv dx^\mu g_{\m\n}dx^\n\, . \label{metric}
 \end{eqnarray}
While the parameterization of the coset only spans one half,
$Z^{n+1}>Z^0$, of the hyperboloid, the physical region within the
intrinsic horizon \be 0>-1+e^{2t} \Vec x^2=\xi^\mu \xi_\mu\,
,\qquad \xi^\mu=(-1,x^i)\, , \label{horizon} \ee is covered by
this coordinate patch.

\section{Arbitrary Spin Bosons}

\label{BOSONS}

Let us now insert matter in our dS (we have only space for
bosons). A massive spin $s$ field  can be described by a
completely symmetric tensor $\varphi_{\mu_1\ldots\mu_s}$ subject
to the field equation and constraints
 \begin{eqnarray}
 \lefteqn{\Big(D_\mu
D^\mu-2n+4+(n-5)s+s^2-m^2\Big)\varphi_{\mu_1\ldots\mu_s}}
\nonumber \\
&&=0 =
D.\varphi_{\mu_2\ldots\mu_s}=\varphi^\rho{}_{\rho\mu_3\ldots\mu_s}\,
. \label{eoms} \end{eqnarray}
 For generic values of the mass $m$,
$\varphi_{\mu_1\ldots\mu_s}$ describes the \be \mu(n,s)\equiv
\frac{(n+2s-2)(n+s-3)!}{s!(n-2)!} \ee degrees of freedom of a spin
$s$ symmetric field in $n+1$ dimensions. The mass parameter is
defined so that the theory is strictly massless for $m^2=0$ with a
gauge invariance
\be\varphi_{\mu_1\ldots\mu_s}=\d_{(\mu_1}\xi_{\mu_2\ldots
\mu_s)}\, , \label{gauge}\ee
 (subject to
$\xi^\rho{}_{\rho\mu_3\ldots\mu_s}=0$). The degree of freedom
count is then
 \be
\mu(n,s)-\mu(n,s\!-\!1)=\frac{(n+2s\!-\!3)(n+s\!-\!4)!}{s!(n\!-\!3)!}
. \label{countfreedom} \ee
 Actions may be written down for these
free theories, both massive and massless.

The physical polarizations of a strictly massless field satisfy
 \be \d.V_{(s-1)}\equiv \d^i V_{ii_2\ldots i_s}=0\, ,
\label{strict} \ee
 thanks to the gauge invariance~\eqn{gauge}
which projects out all but the maximal helicity $s$ excitations. A
field obeying~\eqn{strict} has the correct degree of freedom
count, as given in~\eqn{countfreedom}, for a strictly massless
field.

For partially massless fields,  gauge invariances of the form \be
\delta \varphi_{\mu_1\ldots\mu_s}=\D_{(\mu_1\ldots\mu_t}
\varphi_{\mu_{t+1}\ldots\mu_s)}+\cdots \ee imply that the
requirement~\eqn{strict} is relaxed and replaced by \be
\d^{i_1}\cdots \d^{i_t} V_{i_1\ldots i_s}=0\, , \qquad (t\leq s).
\label{partial} \ee We call such a field ``partially massless of
depth $t$''. This amounts to projecting out all helicities save
$(s,\ldots,t+1)$ and gives $\mu(n,s)-\mu(n,s-t)$ degrees of
freedom.

The subalgebra of translations, dilations and rotations leaves the
condition~\eqn{partial} invariant. However, conformal boosts do
not, unless one tunes the conformal weights $\Delta_s$
appropriately. To obtain these tunings we study \be
\d^{i_1}\cdots\d^{i_t}K_iV_{i_1\ldots i_s}=0\, , \label{condition}
\ee for depth $t$ partially massless polarizations $V_{(s)}$
subject to~\eqn{partial}. It is a simple combinatorics problem to
compute the (unique) value of $\Delta_s$ as a function of the
depth $t$ such that the condition~\eqn{condition} holds. We state
the result below. The main idea is conveyed by the simplest
non-trivial example, spin 2.

For a spin 2 field $V_{ij}-V_{ji}=0=V_i{}^i$, the conformal boost
acts as
 \begin{eqnarray}
 iK_iV_{jk} & = & i(-2y_iD+\Vec y^2 P_i)V_{jk}\nonumber \\
 &+ & 4y_{(j}V_{k)i}
-\delta_{i(j}y_lV_{k)l}\, .
 \end{eqnarray}
 The field $V_{ij}$ is strictly
massless whenever \be \d^iV_{ij}=0\, , \label{div} \ee so we test
whether this condition is respected by conformal boosts by
computing \be \d^k\, K_iV_{jk}=2i(\Delta_s-n)V_{ij}\, . \ee Here
we have relied on the divergence constraint ~\eqn{div}. Hence we
find the strictly massless tuning \be \Delta_s=n\,
.\label{stricttune} \ee Using this relation as well as $s=2$ gives
$m^2=0$, the correct tuning for a strictly massless spin~2 boson.

To study partially massless spin~2, we replace the single
divergence condition~\eqn{div} by the double divergence one \be
\d^{ij}V_{ij}=0.\label{divdiv} \ee Now we must compute \bea
\d^{j}\d^k\, K_iV_{jk}&=&4i(\Delta_s-n+1)\d_jV_{ij} \eea where we
used~\eqn{divdiv} but {\it not}~\eqn{div}. Therefore we obtain the
partially massless spin~2 tuning \be \Delta_s=n-1\,
.\label{partune} \ee It is also clear that the partially massless
representation is irreducible. One might have thought it to be a
direct sum of spin~2 and spin~1 strictly massless representations.
However, since the tunings~\eqn{stricttune} and~\eqn{partune}
differ, the strictly massless spin~2 field components
satisfying~\eqn{div} mix with the remaining ones when
$\Delta_s\neq n$.

We calculate the tunings for arbitrary spin in the same way by
imposing~\eqn{partial} and computing
 \begin{eqnarray}
\lefteqn{\d^{i_1}\cdots\d^{i_t} K_i V_{i_1\ldots i_s}}
\\
&& = \; 2it(\Delta_s-n-s+t+1)\,
\d^{i_2}\cdots\d^{i_{t}}V_{ii_2\ldots i_s}\, .
\nonumber\end{eqnarray}
 The tunings are therefore
 \be \Delta_s=n+s-t-1\, .\label{loony_tunes} \ee
Inserting the tuning condition~\eqn{loony_tunes} in the~dS
mass--conformal weight relation yields the mass conditions for
depth $t$ partial masslessness \be m^2=(t-1)(2s-3+n-t)\, . \ee
 Note that for depth $t=1$,
{\it i.e.} strictly massless fields, the mass parameter $m^2=0$.
When $n=3$, the result agrees with requiring light cone
propagation for all the helicities of partially massless fields.
The above results may be summarized by Figure 1, where the full
$(m^2,\L )$ phase plane for both bosons and fermions is displayed.

\section{Spin 2}

We give here a concrete example of the behavior of a model at
critical tuned values of $m^2/\L$, namely the (simplest) $s$=2
case in $D$=4. After some work, the action decomposes into a
non-interacting sum of helicities, in terms of the standard flat
space orthogonal decomposition of a symmetric tensor $T_{ij}$,
 \be
 I = I_{\pm2}(T_{ij}^{TT}) + I_{\pm 1}(T_i^T) + I_0(T^t,T^l)\ ,
 \label{cat}
 \ee
where the subscripts refer to the helicities covered. Let us begin
with the helicity $\pm$2 part, where there is never a constraint:
 \begin{eqnarray}
 \lefteqn{I_{\pm2} =  p_{\pm2}\, \dot q_{\pm2} -\,\Big[\ \frac{1}{2}\,
 p_{\pm2}^2 \ + \  \frac{1}{2}\, q_{\pm2}\,          }\\
&& ~~~     \Big(-\nabla^2+m^2
                  -\frac{9M^2}{4}\Big)\,q_{\pm2}\ \Big] \
                  .\nonumber
 \label{pm2}
 \end{eqnarray}
 [Here and throughout, we omit all explicit integrals.]
We will explain and meet again the effective mass $(m^2-9M^2/4)$
later, and at present just state that this action ensures stable,
unitary propagation for {\it all} $m^2\geq 0$. Therefore the
helicity $\pm2$ modes propagate according to~\eqn{pm2} for all
models in the $(m^2,\L)$ half-plane; here $M^2 \equiv \Lambda /3$.

The helicity $\pm1$ action is identical to its $\pm2$ counterpart
with one important difference: The field redefinition needed to
reach this form is singular at $m^2=0$, as it should be, since
$I_\pm$, disappears at $m^2=0$.

The physical helicity 0 state leads a more interesting life, as it
can be (i) stable and unitary when $m^2>2M^2$, (ii) absent when
$m^2=2M^2$ or (iii) unstable and nonunitary for $m^2<2M^2$.  After
a great deal of work the full helicity 0 action becomes
 \begin{eqnarray}
  I_0 &=& \!p_0\,\dot q_0 -
\left[\
    -\ \frac{3\ \nu^2\ m^2}{32M^2}\ (h^t)^2
    \ -\ \frac{1}{2M}\, h^t \right. \nonumber \\
  &&  \Big(\ \nabla^2\  \Big[p_0-Mq_0\Big]
    + \frac{\nu^2\ m^2}{4M}\, q_0\ \Big)
\nonumber\\
& - & \frac{2}{m^2}\, \Big[p_0-Mq_0\Big] \ \nabla^2 \
\Big[p_0-Mq_0\Big] \nonumber \\
& + & \left. \frac{3}{2}\, \Big[p_0-Mq_0\Big]\,\Big(
p_0-\frac{m^2}{3M}\, q_0\Big)\: \right]\ ,\nn \\
 \n^2 & \equiv & (m^2 - 2M^2)\; .
\label{full}
\end{eqnarray}
The denominators $M$ in this expression do not represent a genuine
singularity, but arise from choosing to solve the constraint in
terms of $p^t$. In contrast, the denominators $m^2$ are due to
integrating out the shift $N_i$ and are a reminder (as we have
seen already) of the strictly massless $m^2=0$ gauge theory. The
key point is to notice that the coefficient of $(h^t)^2$ vanishes
on the critical line $\nu^2=0$ (as well as at $m^2=0$, concordant
with the previous remark). At criticality, the field $h^t$ appears
only linearly and is a Lagrange multiplier for a new constraint,
whereas for $\nu^2\neq0$, we can integrate out $h^t$ by its
algebraic field equation and there are no further constraints. Let
us deal with each of these cases in turn.

Consider the models with mass tuned to the cosmological constant
via $m^2=2\L/3$. As is clear from~\eqn{full}, the Lagrange
multiplier $h^t$ imposes the new constraint $p_0-M q_0=0$. 
Eliminating $q_0$ (say) and since $(p_0 \dot p_0)$ is a total
derivative, the 0 helicity action~\eqn{full} vanishes exactly,$
I_0=0$ .It is known~\cite{DeserNepomechi} that the critical
theory possesses a local scalar gauge invariance,
 \be \delta h_{\mu\nu}
= (D_{(\m}D_{\n)} + \frac{\L}{3}\,\ol  g_{\mu\nu} )\, \xi(x)\, .
\ee Thus, our result establishes that its effect is to remove the
lowest helicity excitation. Therefore, the total action is
 \begin{eqnarray}
 \lefteqn{I_{\nu^2=0} =  \sum_{\e=(\pm2,\pm1)}\, \left\{ p_{\e}\, \dot q_{\e}
           -\,\Big[\ \frac{1}{2}\, p_{\e}^2 \ + \
                   \frac{1}{2}\, q_{\e}\right.} \nonumber \\
&& ~~~~~ \left. \Big(-\nabla^2
                  -\frac{M^2}{4}\ \Big)\,q_{\e}\ \Big]\right\} \ .
 \end{eqnarray}
[The effective mass $-M^2/4$ is the same one as in (27), evaluated
at $\nu^2=0$.] These remaining helicity $(\pm2,\pm1)$ excitations
are both unitary and, as we will show, stable.

We may now eliminate $h^t$ by its algebraic field equation \be
 h^t     \ =\ -\frac{8M}{3\ \nu^2m^2}\,
     \nabla^2\  \Big[p_0-Mq_0\Big]
    - \frac{2}{3}\, q_0  \ ,
\ee
 and made a penultimate field redefinition/canonical
transformation to obtain
  \begin{eqnarray}
  \lefteqn{I_{0} =  p_{0}\, \dot q_{0}
           -\,\left[\ \frac{1}{2}\,
\left[\!\frac{\nu^2\ m^2}{12M^2}\!\right]\, q_{0}^2 \ + \
                   \frac{1}{2}\,
\left[\!\frac{12M^2}{\nu^2\ m^2}\!\right]\, p_{0}\right.}\nonumber \\
&&~  \Big(-\nabla^2+m^2 \left.
                  -\frac{9M^2}{4}\Big)\,p_{0}\ \right] \ .
\label{pen} \end{eqnarray}
 Before we present the final, complete, action,
some important comments on its penultimate form~\eqn{pen} are
needed:
\begin{itemize}
\item The sign of the parameter $\nu^2$ controls the positivity
of the Hamiltonian (and consequently the energy). Therefore we
find that the $(m^2,\L)$ plane is divided into a stable region
$m^2\geq2\L/3$ and an unstable one $m^2<2\L/3$.  We have already
dealt with the transition line $m^2 = 2\Lambda /3$.
\item A final field redefinition,
\be
 p_0\rightarrow -\ \frac{\nu m}{2\sqrt{3}\ M}\ q_0  ,
\;\; q_0\rightarrow \frac{2\sqrt{3}\ M}{\nu m}\ p_0 \label{last}
 \ee
 brings the helicity 0 action into the same form
as its helicity $(\pm2,\pm1)$ counterparts (27), but this is only
legal in the stable massive region $m^2>2\L/3$.
\item The apparent singularity at $M=0$ is spurious and reflects our
(arbitrary) choice of solution to the constraint.
\end{itemize}
The final action for massive spin~2 in the region $m^2>2\L/3$ is
 \begin{eqnarray}
 \lefteqn{I_{\nu^2>0} =  \sum_{\e=(\pm2,\pm1,0)}\, \left\{ p_{\e}\, \dot q_{\e}
           -\,\Big[\ \frac{1}{2}\, p_{\e}^2 \ + \
                   \frac{1}{2}\, q_{\e} \right. }\nonumber \\
 && ~~~~\left. \Big(-\nabla^2 +m^2
                  -\frac{9M^2}{4}\ \Big)\,q_{\e}\ \Big]\right\} \ ,
 \end{eqnarray}
and describes $2s+1 = 5$ stable, unitary, helicity $(\pm2,\pm1,0)$
excitations.

We are now ready to demonstrate the stability of the model in the
allowed region $m^2\geq 2\L/3$. The dS background possesses a
Killing vector \be \ol\xi^\m=(-1,Mx^i)\, , \qquad
\ol\xi^2=-1+\Big(fMx^i\Big)^2 \, , \ee timelike within the
intrinsic horizon $(fMx^i)^2=1$. Therefore, in this region of
spacetime, it is possible to define a conserved energy whose
positivity guarantees the stability of the model.

Let us consider  helicity $\e$ (omitting 0 at criticality)
described by the conjugate pair $(p_\e,q_\e)$, whose time
evolution is generated by the Hamiltonian
 \be H_\e=\ \frac{1}{2}\,
p_{\e}^2 + \frac{1}{2}\, q_{\e}\, \Big(-\nabla^2+m^2
 -\frac{9M^2}{4}\ \Big)\,q_{\e} .
\label{pig} \ee
  However, $H_\epsilon$ is not
conserved, thanks to the explicit time dependence $f^{-2}(t)$ in
$\nabla^2$, which was to be expected since it generates time
evolution $\frac{d}{d t}$ rather than along the Killing direction
$\ol\xi^\m\d_\m$. Instead, the conserved energy is defined in
terms of the stress tensor
 \be E_\e=T_\e^0{}_\m \ {\ol\xi}^\m = -
T_\e{}^0{}_0
 + M\  x^i\  T_\e{}^0{}_i\, .
\ee The momentum density $T_\e{}^0{}_i$ will be defined below and
$-T_\e{}^0{}_0=H_\e$. For gravity, the momentum density $T^0_i$ is
the quadratic part of the coefficient of $N_i$, and a similar
result holds here. Keeping track of our field redefinitions, we
find (modulo irrelevant superpotentials)
 \be
 T_\e{}^0{}_i = - p_\e\ \pa_i\  q_\e
+ \frac{1}{2} \: \pa_i \ \Big(p_\e\, q_\e\Big)\, .
 \ee
It is not difficult to verify that the energy function \be
E_\e=H_\e-M\ x^i\ p_\e \pa_i  q_\e - \frac{3}{2}\ M \ p_\e q_\e \,
, \ee is indeed conserved, $\dot E=0$.

Finally we come to positivity. Here we need only a simple
extension of a method used in analyzing energy in dS. Rewriting
$E$ as
\begin{eqnarray}
E & = & \frac{1}{2}\,\Big(\wh x^i\,\wt p_\ve\Big)^2
+\frac{1}{2}\,\Big(f^{-1}\d_i\,q_\ve\Big)^2\nonumber \\
& - & fM |x|\,\Big(\wh x^i\,\wt
p_\ve\Big)\,\Big(f^{-1}\,\d_i\,q_\ve\Big) +
\frac{1}{2}\,m^2\,q_\ve^2\, ,\nonumber \\
&& \wt p_\e \equiv p_\e -\frac{3M}{2}\ q_\e\ , \;\;\; x^i\equiv
|x|\,\wh x^i\ , \label{triangle} \end{eqnarray}
 the first three
terms are positive by the triangle equality within the intrinsic
dS horizon \be f\,M \,|x|<1\ , \ee and the fourth, mass term is
manifestly positive\footnote{The Killing energy of a massive
scalar  in dS also takes the form~\eqn{triangle} and is therefore
stable for $m^2\geq0$. [In this non-gauge example, there is no
analog to the spin~2 instabilities at negative values of $m^2$ or
$\nu^2$ whose vanishing is associated with gauge invariances.]
Scalars in AdS actually enjoy a somewhat wider stability range,
extending to negative values of $m^2$ due to a shift in the
spectrum of the AdS 3-Laplacian. This broadening is unlikely for
spin~2, since its stability is controlled entirely by the above
gauge coefficients.} This concludes our stability proof.

The instability of the region $m^2<2\L/3$ is also manifest:
Consider helicity $\e=0$. Recall that once $\nu^2<0$, we cannot
make the rescalings with factors $\nu$ and $\nu^{-1}$ in the final
field redefinition~\eqn{last}. This does not prevent us from
constructing a conserved energy with a ``triangle''
form~\eqn{triangle}, but the {\it caveat} is that  $p_0^2$ carries
a factor $\nu^2$ and likewise $q_0^2$ a factor $\nu^{-2}$.
Therefore the energy is negative and the theory is unstable in
this region, entirely due to helicity 0.

\section{Inconsistencies of massive charged gravitating higher spins}

As a final topic, we rapidly review the problems of higher spin
couplings to gauge and gravity fields, and a shorter way to obtain
them. Localized, massive $s>1$ particles have never been observed,
in agreement with a large (if somewhat confusing) higher spin lore
that they cannot be made to interact consistently even with
gravity or electromagnetism. Here we combine two earlier lines of
analysis into a systematic study of higher spins coupled to
Einstein--Maxwell fields.

(i) Massive higher spins propagate consistently in constant
curvature backgrounds for a range of parameters $(m^2,\Lambda)$
centered around the Minkowski line $(m^2,0)$. (ii) The original
unitarity (or equivalently causality) difficulties of massive
$s=3/2$ persist in pure E/M backgrounds, even including all
possible non-minimal couplings.

Our first new result is that the onset of the unitarity/causality
difficulty for massive $s=3/2$ in pure E/M backgrounds can be
traced to a novel gauge invariance of the timelike component of
the Rarita--Schwinger equation at E/M field strengths tuned to the
mass. Although the full system is not invariant, a consequence of
this invariance is signal propagation with lightlike
characteristics. Beyond this tuned point, {\it i.e.}, for large
enough magnetic field $\vec B^2>(\frac{3m^{2}}{2e})^{\!^{^2}}$ (or
better, small/large enough mass/charge), the system is neither
causal nor unitary. This is an old result \cite{JohnsonSudarshan}
but its rederivation in terms of a gauge invariance is edifying.
Our second $s=3/2$ result is that in {\it dynamical}
Maxwell--Einstein backgrounds, causality can be maintained for any
choice of E/M field, for certain values of the mass.

The first instance in which there is no underlying protection, in
contrast to that provided by SUGRA for $s=3/2$ is $s=2$ is tensor
theory: when charged it bears little resemblance to its one
conceivable relative, Einstein gravity. Charged massive $s=2$
preserves the correct DoF only for gyromagnetic ratio $g=1/2$, but
even this theory suffers from the usual causality difficulties.
Furthermore, it has no good DoF coupling to general gravitational
backgrounds, so there is no analog of the causality bounds found
for $s=3/2$.

We examine the leading discontinuities of a shockwave, which are
now second order and denoted as
 \be
[\d_\m\d_\n\phi_{\r\s}]_{_{\scriptstyle \Sigma}}
=\xi_\m\xi_\n\Phi_{\r\s}\, .
 \ee
 From the field equation $\xi_{\mu\nu}=0$ and its
trace we learn that
 \bea [{\cal G}_{\m\n}-\frac{1}{2}\,
\eta_{\m\n}{\cal G}_\r{}^\r]_{_{\scriptstyle \Sigma}} & \!\! =
\!\! &
\xi^2\Phi_{\m\n}+\xi_\m\xi_\n\Phi \nonumber \\
& \! - \! & 2\,\xi_{(\m}\xi.\Phi_{\n)}=0 ,
\label{maroon}\\
{}[{\cal G}_\m{}^\m]_{_{\scriptstyle \Sigma}} & \! = \!
&-2\xi^2\Phi+2\xi.\xi.\Phi=0  , \label{chatreuse}
 \eea
 ($\Phi\equiv\Phi_\m{}^\m$). We now study the system for a
causal timelike normal vector $\xi^2=-1$. Note that since
$\Phi_{\m\n}\neq0$, we deduce that $V_\m\equiv\xi.\Phi_\m\neq0$
(otherwise $\xi^2=0$ and the model would be causal). So we now
impose
 \be \Phi_{\m\n}=-\xi_\m\xi_\n\,\xi.V-2\,\xi_{(\m}V_{\n)}\,
,~~~ \Phi=-\xi.V \ee
 and study further constraints. In
particular, the single divergence constraint gives
 \begin{eqnarray}
\lefteqn{\xi^\m[\d_\m D.{\cal G}_\n]_{_{\scriptstyle \Sigma}} =
m^2(V_\n+\xi_\n\,\xi.V)+ \frac{3ie}{2}}\nonumber \\
&& \Big( F_{\n\r}V^\r -\xi^\r F_{\r\n}\,\xi.V+\xi_\n\,\xi.F.V
\Big)\, ,
 \end{eqnarray}
  so that
   \be
\Pi_{\n\r}\Big[m^2\eta^{\r\s}+\frac{3ie}{2}\,F^{\r\s}\Big]\,
\Pi_{\s\tau}\,V^\tau=0\, , \label{traceless}
 \ee
 where the
projector $\Pi_{\m\n}\equiv\eta_{\m\n}+\xi_\m\xi_\n$. It is
sufficient to search for acausalities in constant background E/M
fields, so with this restriction the double divergence constraint
gives
 \begin{eqnarray}
 \lefteqn{ \xi^\m\xi^\n\,[\d_\m\d_\n(D.D.{\cal
G}+\frac{1}{2}\,m^2{\cal G}_\r{}^\r)]_{_{\scriptstyle \Sigma}} =
-\frac{3}{2}}\nonumber \\
&& (m^4-\frac{e^2}{2}\,F^2+e^2\,\xi.T.\xi)\, \xi.V
-3e^2\,\xi.T.V\, . \label{trace} \end{eqnarray}
 Causality requires that the
system of equations~\eqn{traceless} and~\eqn{trace} have no
non-zero solution for $V_\m$. In general~\eqn{traceless} implies
the vanishing of the components of $V_\m$ orthogonal to $\xi_\m$
and~\eqn{trace} in turn removes the parallel components. However
if, regarded as a matrix equation in the orthogonal subspace,
equation~\eqn{traceless} fails to remove the orthogonal components
of $V_\m$ the model will be acausal. The determinant in this
subspace vanishes whenever the (long-known) bound \cite{Kobayashi}
 \be \vec
B^2=\left(\!\frac{2m^2}{3e}\!\right)^2\, . \ee
 is reached,
Requiring that~\eqn{trace} be non-degenerate yields a different
and weaker bound $\vec B^2=(2m^4+\vec E^2)/(3e^2)$. These
differing bounds for the propagation of helicities zero and one
are reminiscent of the behaviour found for $s=2$ in cosmological
backgrounds. Needless to say, these acausality and DoF problems
are entirely independent of the formalism used to describe the
initial free fields and to introduce couplings; these can only
differ by non-minimal terms.

\section{dS Supergravity?}

Our considerations thus far have led us to the following picture
of particles in (A)dS backgrounds. Partially massless fields, of
either statistics, are always unitary in~dS, while in AdS only
strictly massless ones are. This behavior is understood by turning
on cosmological constants of either sign and following their
effects on the signs of lower helicity state norms. A sequence of
unitary partially massless fields is only encountered when
starting from Minkowski space ($\Lambda=0$) and first turning on a
positive cosmological constant. The bad news, however, is that
partially massless fermions require tunings with negative $m^2$ as
already noted in cosmological supergravity. This led to the
rejection of dS supergravity as a consistent local QFT, a
rejection bolstered later by the difficulties in defining string
theory on~dS backgrounds.  Here I want to offer a few speculative
remarks (every lecture should have some) revisiting this question.

Let us first present the reasons for rejection in terms of the
present analysis, followed by such mitigating circumstances as we
can muster for keeping the possibility in play. For concreteness,
we deal primarily here with $N=1$ supergravity in four dimensions.

\begin{itemize}
\item Although (formally) locally supersymmetric actions exist, the mass
parameter appearing in the term $m\sqrt{-g}\,
\psi_\m\gamma^{\m\n}\psi_\n$ must be pure imaginary for lightcone
propagation. Therefore, the action of~dS supergravity does not
obey a reality condition.

\item The associated local supersymmetry
transformations $\delta\psi_\mu=(D_\mu+\frac12
m\gamma_\mu)\varepsilon$, being complex, cannot preserve Majorana
condition on fields. For $N=2$ supersymmetry, a symplectic reality
condition is possible, but the locally supersymmetric action is
either still complex or the Maxwell field's kinetic term has
tachyonic sign.

\item Another way to see that there can be
no real supercharges is that the $N=1$~dS superalgebra ought be
the $d=5$, $N=1$ superlorentz algebra, but there are no Majorana
spinors in $d=5$ Minkowski space.

\item Because~dS$_4$ has topology $S^3\times
{\mathbb R}$, gauge charges (being surface integrals) vanish since
there are no spatial boundaries. This argument is of a different
nature from the previous ones as it involves the global
considerations which we have chosen to ignore.

\item Finally, even for the $N=2$ case, where a~dS superalgebra
exists, there are no positive mass, unitary representations.
Unlike the praiseworthy AdS algebra, our maximal compact
subalgebra has no $SO(2)$ factor whose generator could define a
positive mass Hamiltonian. Again this is a global issue.
\end{itemize}

Let us now present the arguments in favor of~dS supergravity:
\begin{itemize}
\item While the tuned~dS supergravity mass $m^2=-\Lambda/3\, .$
is indeed negative, there are precedents for consistent theories
with negative~$m^2$, for example scalars in AdS for which a range
of such values can be tolerated essentially because the lowest
eigenvalue of the Laplacian has a positive offset there. Fermions
mirror this behavior in~dS.

\item Despite an imaginary
mass-like term in the action, at least the free field
representations are unitary. For unitarity of spin~3/2
representations, the relevant quantity is $m^2-3\Lambda$, not
$m^2$ alone. In addition the linearized equations of motion for
physical spin~2, helicity~$\pm2$ and its proposed spin~$3/2$,
helicity~$\pm3/2$ superpartner degrees of freedom are identical.

\item In~dS space positivity of energy is possible only for
localized excitations within the horizon. Only this region of~dS
possesses a timelike Killing vector.~dS gravity is therefore
stable against local excitations within the physical region.
Although~dS quantum gravity is problematic, nobody would reject
cosmological Einstein gravity as an effective description of local
physics within the physical region. This looser criterion is all
we should require of a sensible~dS supergravity.
\item
As stated, any local supersymmetry is purely formal in the absence
of $d=4+1$ Majorana spinors (even the fermionic field equation
above is not consistent with a reality condition on $\psi_\mu$).
In any case, a bona fide $N=1$~dS superalgebra with Majorana super
charge, would imply a global positive energy theorem from
$\{Q,Q\}\sim H$, which is already ruled out at the level of~dS
gravity. Instead, we could envisage relaxing the Majorana
condition and allowing only a formal local supersymmetry. A direct
Hamiltonian constraint analysis still shows that only
helicities~$\pm3/2$ propagate. Also there are other examples where
the Majorana condition must be dropped, but nonetheless a formal
supersymmetry yields the desired Ward identities: continuation to
Euclidean space is a familiar case.
\end{itemize}

The summary in favor of~dS supergravity is then that the free
field limit is unitary with time evolution governed by the
Hermitean generator $D=iM_{40}$ subject to a positive energy
condition--within the physical region inside the Killing horizon.
A formal supersymmetry, similar to that remaining when
Euclideanizing supersymmetric Minkowski models, suffices to show
that amplitudes obey supersymmetric Ward identities.

\section{Acknowledgements}

I thank the organizers, especially Ilya Shapiro, whose dedication
made this school successful. It is also a pleasure to thank Andrew
Waldron for suggestions on this version of our work. This research
was supported by NSF grant PHY99-73935.

\newpage
\vskip 2 cm 
\begin{figure}
\epsfxsize=1.5cm
\centerline{
\epsffile[5 30 100 150]{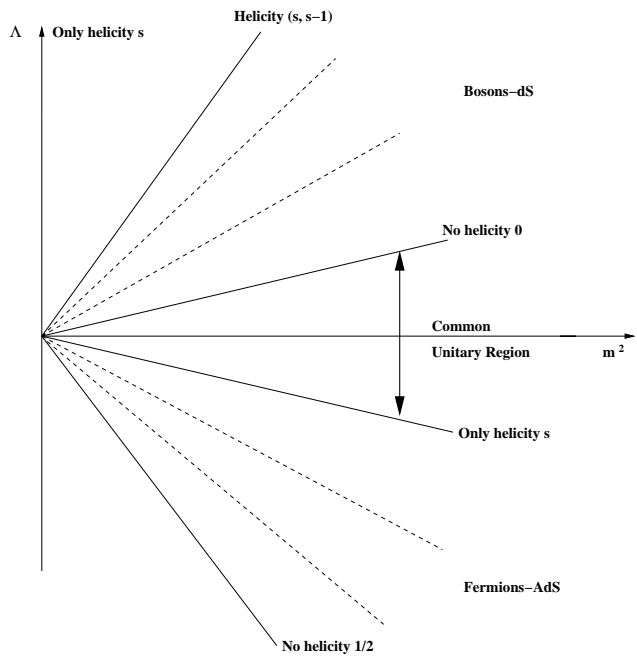}}
\caption{ The top/bottom halves of
the half-plane represent dS/AdS (and also bosons/fermions)
respectively. The $m^2=0$ vertical is the familiar massless
helicity $\pm s$ system, while the other lines in dS represent
truncated (bosonic) multiplets of partial gauge invariance: the
lowest has no helicity zero, the next no helicities $(0,\pm 1)$,
etc.  Apart from these discrete lines, bosonic unitarity is
preserved only in the region below the lowest line, namely that
including flat space (the horizontal) and all of AdS. In the AdS
sector, it is the topmost line that represents the pure gauge
helicity $\pm s$ fermion, while the whole region below it,
including the partially massless lines, is non-unitary. Thus, for
fermions, only the region above the top line, including the flat
space horizontal and all of dS, is allowed.  Hence the overlap
between permitted regions straddles the $\L = 0$ horizontal and
shrinks down to it as the spins in the tower of spinning particles
grow; only $\L = 0$ is allowed for generic ($m^2$ not growing as
$s^2$) infinite towers.}
\label{fig:1}
\end{figure}

\end{document}